\documentclass[journal=jctcce,manuscript=article,layout=onecolumn]{achemso}

\usepackage[version=3]{mhchem} 
\usepackage{subcaption} 
\usepackage{xcolor} 
\usepackage{listings} 
\usepackage{multirow} 
\usepackage[utf8]{inputenc}
\usepackage[pagewise]{lineno}
\definecolor{backcolour}{rgb}{0.95,0.95,0.92}
\DeclareUnicodeCharacter{2212}{-}

\usepackage{xr} 
\usepackage[scaled]{helvet}
\usepackage[T1]{fontenc}

\lstdefinestyle{mystyle}{
    backgroundcolor=\color{backcolour},
    commentstyle=\color{red},
    keywordstyle=\color{purple},
    numberstyle=\tiny\color{gray},
    stringstyle=\color{red},
    basicstyle=\tiny, 
    breakatwhitespace=false,
    breaklines=true,
    captionpos=b,
    keepspaces=true,
    numbers=left,
    numbersep=5pt,
    showspaces=false,
    showstringspaces=false,
    showtabs=false,
    tabsize=2,
    aboveskip=3mm,
    belowskip=3mm
}
\usepackage{tikz} 

\author{Sung Wook Moon}
\affiliation[UNIST]
{Department of Chemistry, School of Natural Science, Ulsan National Institute of Science and Technology (UNIST), 50 UNIST-gil, Ulju-gun, Ulsan 44919, South Korea}
\author{Soohaeng Yoo Willow}
\affiliation[SKKU]
{Department of Energy Science, Sungkyunkwan University, Seobu-ro 2066, Suwon, 16419, Korea
}
\author{Tae Hyeon Park}
\affiliation[SKKU]
{Department of Energy Science, Sungkyunkwan University, Seobu-ro 2066, Suwon, 16419, Korea
}
\author{Seung Kyu Min}
\affiliation[UNIST]
{Department of Chemistry, School of Natural Science, Ulsan National Institute of Science and Technology (UNIST), 50 UNIST-gil, Ulju-gun, Ulsan 44919, South Korea}
\alsoaffiliation[IBS]
{Center for 2D Quantum Heterostructures, Institute for Basic Science (IBS), Suwon 16419, Korea
}
\email{skmin@unist.ac.kr}
\author{Chang Woo Myung}
\affiliation[SKKU]
{Department of Energy Science, Sungkyunkwan University, Seobu-ro 2066, Suwon, 16419, Korea
}
\alsoaffiliation[IBS]
{Center for 2D Quantum Heterostructures, Institute for Basic Science (IBS), Suwon 16419, Korea
}
\email{cwmyung@skku.edu}

\title
  {Machine Learning Nonadiabatic Dynamics: Eliminating Phase Freedom of Nonadiabatic Couplings with the State-Interaction State-Averaged Spin-Restricted Ensemble-Referenced Kohn-Sham Approach}

\begin{document}
\begin{tocentry}

%
%
%
\centering
\includegraphics[scale=0.25]{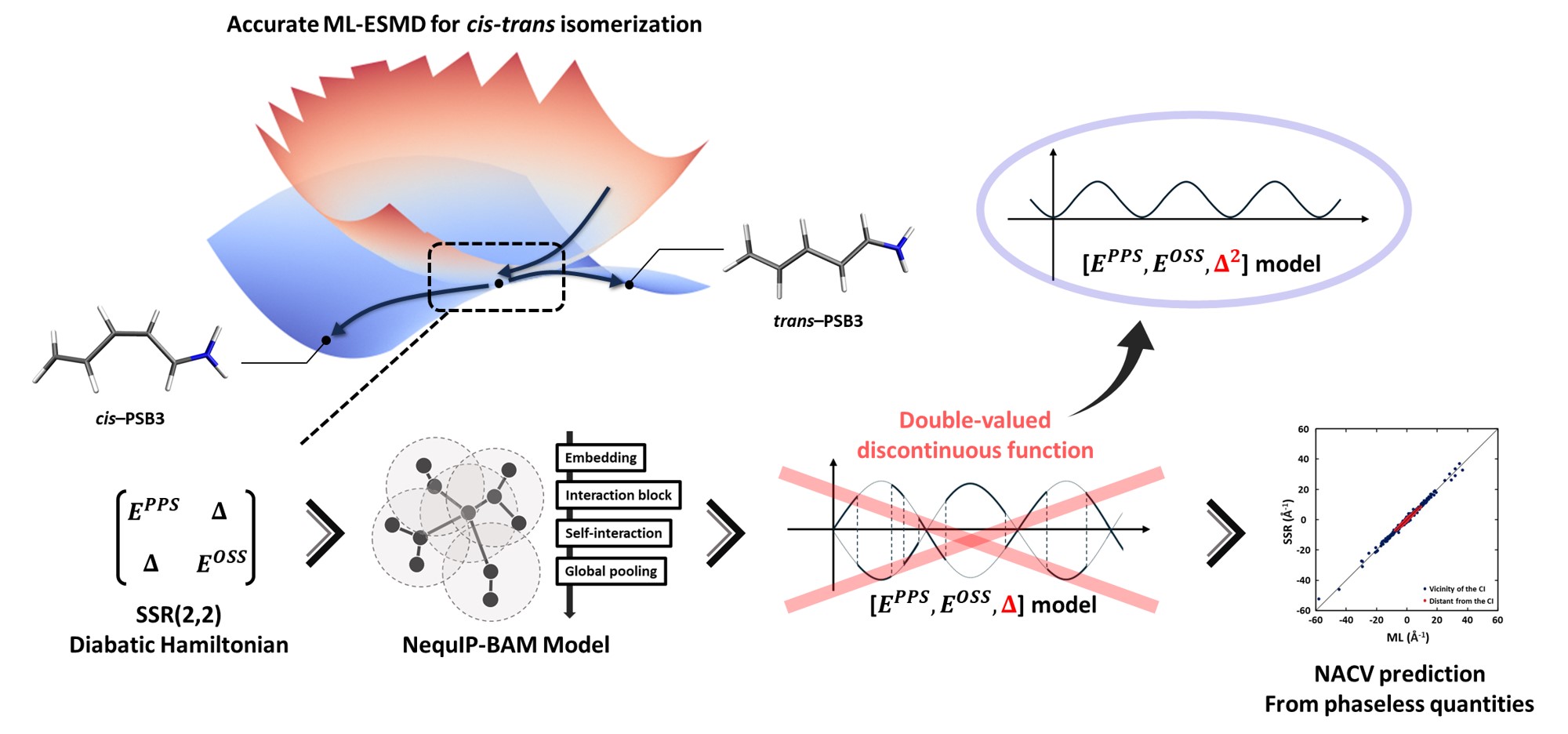}

\end{tocentry}


\begin{abstract}
Excited-state molecular dynamics (ESMD) simulations near conical intersections (CIs) pose significant challenges when using machine learning potentials (MLPs). Although MLPs have gained recognition for their integration into mixed quantum-classical (MQC) methods, such as trajectory surface hopping (TSH), and their capacity to model correlated electron-nuclear dynamics efficiently, difficulties persist in managing nonadiabatic dynamics. Specifically, singularities at CIs and double-valued coupling elements result in discontinuities that disrupt the smoothness of predictive functions. Partial solutions have been provided by learning diabatic Hamiltonians with phaseless loss functions to these challenges. However, a definitive method for addressing the discontinuities caused by CIs and double-valued coupling elements has yet to be developed. Here, we introduce the phaseless coupling term, $\Delta^2$, derived from the square of the off-diagonal elements of the diabatic Hamiltonian in the state-interaction state-averaged spin-restricted ensemble-referenced Kohn-Sham (SI-SA-REKS, briefly SSR)(2,2) formalism. This approach improves the stability and accuracy of the MLP model by addressing the issues arising from CI singularities and double-valued coupling functions. We apply this method to the penta-2,4-dieniminium cation (PSB3), demonstrating its effectiveness in improving MLP training for ML-based nonadiabatic dynamics. Our results show that the $\Delta^2$ based ML-ESMD method can reproduce ab initio ESMD simulations, underscoring its potential and efficiency for broader applications, particularly in large-scale and long-timescale ESMD simulations.
\end{abstract}


\section{Introduction}\label{sec:intro}
The study of excited state phenomena has been a significant focus for researchers across various fields, offering profound insights into processes, such as photoisomerization\cite{bandara_2012, kucharski_2014, piermichele_2023, pieroni_2024}, photocatalysis\cite{hoffmann_2008, keun_2021, kim_2021}, solar cells\cite{blancon_2017, dey_2021, edoardo_2022}, and many other light-involved chemical processes\cite{pdt_2003, fraga_2008, coskun_2012, anderson_2021, peng_liu_2023, bhuyan_2023}. In computational chemistry, the light-matter interaction in molecular systems can be explored using excited state molecular dynamics (ESMD). ESMD provides chemical insights by allowing for time-resolved tracking of molecular systems. Unlike ground state dynamics, where the nuclear wavepacket fluctuation is confined to a single adiabatic potential energy surface (PES), ESMD considers the presence of nonadiabatic interactions across multiple PESs.

From a practical perspective, mixed quantum-classical (MQC) approaches\cite{TSH_1990, ehrenfest_1999, deco_prezhdo_2012, SCMC_2013, jkha_deco_2018} are commonly used where MQC methods treat electronic propagation quantum mechanically while the nuclei move classically along multiple trajectories. Since conventional photochemical reactions occur over timescales ranging from hundreds of femtoseconds to nanoseconds, a large number of electronic structure calculations are required during an ESMD simulation. For instance, predicting PESs for 100,000 nuclear configurations is necessary to simulate dynamics for 1 ps with a time step size of 1 fs and 100 nuclear trajectories. Moreover, ESMD is highly sensitive to the accuracy of nonadiabatic coupling vectors (NACVs), particularly in cases involving conical intersections (CIs). In such scenarios, quantum chemical methods that adequately account for electronic correlation must be chosen. Although there are highly accurate quantum mechanical methods, e.g. equation-of-motion coupled-cluster singles and doubles method and its derivatives\cite{cc_1993, cc_2007, kjønstad_2017, gulania_2021}, second-order perturbation theory with a complete active space self-consistent field\cite{Andersson1990}, and density matrix renormalization group\cite{Hu2015}, MQC simulations require hundreds or even thousands of trajectories, making it computationally prohibitive to use such high-level ab initio theories, even for small systems. This makes the trade-off between accuracy and computational cost a central issue in MQC dynamics, presenting an ongoing challenge in balancing accuracy with feasibility.

Over the past decade, machine learning potentials (MLPs) have enabled us to overcome the accuracy-cost trade-off in theoretical chemistry. MLPs can predict molecular properties with an accuracy comparable to traditional ab initio theories but at a significantly lower computational cost. \cite{unkeMachineLearningForce2021,DeringerCsanyi21,BartokCsanyi10,ChenMarkland23a,gilmerNeuralMessagePassing2017,batznerEquivariantGraphNeural_2022,hajibabaeiSparseGaussianProcess2021,hajibabaeiUniversalMachineLearning2021,HajibabaeiKim21b,myungChallengesOpportunitiesProspects2022,VandermauseKozinsky22,VandermauseKozinsky20,metcalfApproachesMachineLearning2020,naReverseGraphSelfattention2021,zhung3DMolecularGenerative2024,imbalzanoUncertaintyEstimationMolecular2021,Bayerl2022,bogojeski_2020, zuo_2020, willow2024, dral_barbatti_2018, na_chang_kim_2020, schran_2021} These advances in ML-based property prediction can be integrated into the MQC dynamics by replacing quantum mechanical calculations with MLPs. However, applying MLPs to ESMD introduces unique challenges that are not typically encountered in ground state MD simulations. A key difficulty is the prediction of NACVs near CIs of PESs. The NACV formula incorporates the energy difference between two electronic states in the denominator. Even minor energy errors can cause significant NACV inaccuracies near CIs. Additionally, NACVs are not differentiable and diverge at CIs. Various studies have proposed solutions to the singularities at CIs. The first approach is to avoid the direct prediction of NACV by using a method that does not require NACV. The simplicity of the Landau–Zener formalism\cite{belyaev_2014} has attracted researchers and has been employed in various studies\cite{pios_dral_2024, zhang_barbatti_2024}.
The Zhu-Nakamura method\cite{zhu_nakamura_1994}, based on the Landau–Zener formalism, has also been utilized in ML-based noadiabatic dyanmics\cite{da_2018, chen_dral_2018}. Another approach focused on learning diabatic Hamiltonian instead of adiabatic Hamiltonian\cite{shu_2020, axelrod_2022, li_2023}. This provided smooth and differentiable predictions for energy and NACVs near CIs.

While the issue of CI singularity can be addressed through diabatic transformation, another challenge arises from the double-valued nature of the coupling elements. This double-valued nature, known as the geometric phase effect\cite{ryabinkin_2017},  imparts a similar double-valued character to NACVs. As a result, NACVs derived from electronic structure calculations have arbitrary signs, leading to discontinuities that hinder efficient model training. 
To circumvent this problem, a phaseless loss function was introduced\cite{SchnetSharc_2020} and applied in several authors of this work\cite{ESNAMD_2021}. However, this approach does not fully resolve the problem, as the training set contains two values (one positive and one negative) for the same molecular geometry. 
Recent advancements in MLPs have introduced various strategies to address the sign problem in ESMD simulations. For adiabatic states, MLPs have been trained to directly predict adiabatic energies while addressing nonadiabatic coupling vectors through techniques such as phaseless loss functions\cite{SchnetSharc_2020} or auxiliary single-valued functions\cite{jeremy_2023}, which circumvent the sign ambiguity inherent in nonadiabatic couplings. In diabatic states, where the sign problem manifests in the off-diagonal elements of the Hamiltonian, similar approaches have been employed, including phaseless loss functions\cite{ESNAMD_2021} to mitigate this issue.
Our approach differs by directly training MLPs on phaseless quantities within the diabatic framework, ensuring that the training process avoids the complexities introduced by phase dependence. By leveraging the inherent strengths of MLPs, which utilize both function values and their derivatives for higher accuracy, we can efficiently learn diabatic Hamiltonian matrix elements while avoiding the challenges associated with formulating non-adiabatic coupling vectors in adiabatic states. 


Here, we introduce a state-interaction state-averaged spin-restricted ensemble-referenced Kohn-Sham (SI-SA-REKS, briefly SSR)(2,2)\cite{SSR_2008, SSR_2013, SSR_2014} formalism in combination with training of phaseless quantities to address both the CI singularities and the double value problem of coupling elements. Our approach is based on deriving the SSR(2,2) equations using the phaseless coupling, $\Delta^2$, which represents the square of the off-diagonal elements in the diabatic Hamiltonian, rather than the phase-dependent $\Delta$. We trained equivariant message-passing MLPs to model diabatic PESs and the phaseless $\Delta^2$ for the penta-2,4-dieniminium cation (PSB3). This method resolves the phase ambiguity of coupling elements and enhances the accuracy of MLP prediction of NACV in SSR(2,2) ESMD simulations.

\section{Method}\label{sec:method}
\subsection{SSR(2,2) methodology}
We employ the SSR(2,2) methodology to calculate excited electronic states. In the SSR(2,2) method, an ensemble of ground and excited microstates provides $(2 \times 2)$ diabatic Hamiltonian, incorporating both perfectly spin-paired singlet (PPS) and open-shell singlet (OSS) configurations. These configurations are determined by two active Kohn-Sham (KS) orbitals, $\phi_a$ and $\phi_b$, along with their fractional occupation numbers (FONs), $n_a$ and $n_b$. From these orbitals, a total of six microstates can be configured, enabling the construction of both PPS and OSS configurations as

\begin{equation}
    E^{PPS} = \sum\limits_{L=1}^6C_L^{PPS}E_L
\end{equation}
where $C_1^{PPS}=n_a/2$, $C_2^{PPS}=n_b/2$, $-C_3^{PPS}=-C_4^{PPS}=C_5^{PPS}=C_6^{PPS}=f(n_a, n_b)$ and
\begin{equation}
    E^{OSS} = \sum\limits_{L=3}^6C_L^{OSS}E_L
\end{equation}
where $C_3^{OSS}=C_4^{OSS}=1$, $C_5^{OSS}=C_6^{OSS}=-1/2$ with KS energy of six microstates, $E_1=E^{KS}[\phi_a\overline{\phi}_a]$, $E_2=E^{KS}[\phi_b\overline{\phi}_b]$, $E_3=E^{KS}[\phi_a\overline{\phi}_b]$, $E_4=E^{KS}[\overline{\phi}_a\phi_b]$, $E_5=E^{KS}[\phi_a\phi_b]$, and $E_6=E^{KS}[\overline{\phi}_a\overline{\phi}_b]$, respectively.
$f(n_a, n_b)$ is an interpolating function \newline$f(n_a, n_b)=(n_an_b)^{(1-(1/2)((n_an_b+\delta)/(1+\delta)))}$ with $\delta=$0.4.
State-averaged energy functional $E^{SA}$ is constructed from $E^{PPS}$ and $E^{OSS}$, by $E^{SA}=w_{PPS}E^{PPS}+w_{OSS}E^{OSS}$. Here, we choose equiensemble condition, $w_{PPS}=w_{OSS}=1/2$.
From the optimization of orbitals, we obtain the pseudo-Fock equation,
\begin{equation}
n_p\hat{F}_p\phi_p=\sum\limits_{q}\epsilon^{SA}_{pq}\phi_q
\end{equation}
where Fock operator $\hat{F}_p$ is given by
\begin{equation}
\hat{F}_p=\sum\limits_{L}C^{SA}_L {\frac{n^L_{p\alpha}\hat{F}^L_\alpha + n^L_{p\beta}\hat{F}^L_\beta}{n_p}}.
\end{equation}
Here, $n_p$ is an averaged occupation $n_p=\sum\limits_{L}C_L^{SA}(n^L_{p\alpha}+n^L_{p\beta})$, $n_{p\alpha/\beta}^{L}$ is an occupation number of the $p$th orbital in the $L$the microstate configuration, $C_L^{SA}$ is an averaged weight $C_L^{SA}=w_{PPS}C_L^{PPS}+w_{OSS}C_L^{OSS}$, and $\epsilon^{SA}_{pq}$ is a Lagrange matrix element. 
Finally, we can construct $(2 \times 2)$ secular equation to obtain ground and first excited adiabatic state energies, $E_-^{SSR}$ and $E_+^{SSR}$, as

\begin{equation}
    \begin{pmatrix}
    E^{PPS} & \Delta^{SA} \\
    \Delta^{SA} & E^{OSS} \\
    \end{pmatrix}
    \begin{pmatrix}
    a_{00} & a_{01} \\
    a_{10} & a_{11} \\
    \end{pmatrix}
    =
    \begin{pmatrix}
    E_{-}^{SSR} & 0 \\
    0 & E_{+}^{SSR} \\
    \end{pmatrix}
    \begin{pmatrix}
    a_{00} & a_{01} \\
    a_{10} & a_{11} \\
    \end{pmatrix}  
\end{equation}
where we focus on real-valued $\Delta^{SA}$ since conventional quantum chemistry program provides real-valued coupling terms.

The state interaction between PPS and OSS configuration, $\Delta^{SA}$, is calculated from a Lagrange matrix element between two active orbitals $\phi_a$ and $\phi_b$, and FONs as

\begin{equation}
\Delta^{SA}=(\sqrt{n_a}-\sqrt{n_b})\epsilon^{SA}_{ab}
\end{equation}
where FONs $n_a$ and $n_b$ can be obtained from the optimization of KS orbitals while minimizing $E^{SA}$.
In the SSR(2,2) method, the NACV between two adiabatic states can be directly calculated from the diabatic elements $E^{PPS}$,$ E^{OSS}$, and $\Delta^{SA}$ by
\begin{equation}
    d_{01} = \frac{1}{E_+^{SSR} - E_-^{SSR}}((a_{00}a_{01} - a_{10}a_{11})\vec{g} + (a_{00}a_{11} + a_{01}a_{10})\vec{h})
\end{equation}
where $\vec{g} = \frac{1}{2}\nabla(E^{PPS}-E^{OSS})$ and $\vec{h} = \nabla\Delta^{SA}$ while the coefficients $a_{ij}$ can be expressed with $E^{PPS}$, $E^{OSS}$, and $\Delta^{SA}$ by solving the secular equation. 

\subsection{Adiabatic energies and NACVs from phaseless quantities}
Our key observation is that the NACV of Equation (7) can be rewritten based on the phaseless quantity $\Delta^{2}$ (where $\Delta=\Delta^{SA}$, for simplicity) as follows:

\begin{equation}\label{eq:nacv}
    d_{01} = \frac{nm}{E_+^{SSR} - E_-^{SSR}}((-\Delta^2)\nabla(E^{PPS}-E^{OSS}) + \frac{1}{2}(E^{PPS}-E^{OSS})\nabla\Delta^{2})
\end{equation}
where $n$ and $m$ represent normalized constants, which have a value of \newline$n=\pm1/\sqrt{(E_- - E^{OSS})^2 + \Delta^2}$ and $m=\pm1/\sqrt{(E_+ - E^{OSS})^2 + \Delta^2}$, respectively (See the derivation in Supporting Information S1).
The presence of the $\pm$ sign in $n$ and $m$ indicates the phase of the NACV, which arises from the phase dependency of the eigenvector in Equation (5).
The significance of Eq. \ref{eq:nacv} lies in the complete substitution of $\Delta$ with $\Delta^2$.
By this substitution, instead of learning the NACV affected by phase arbitrariness, we focus on learning the components of the NACV that are free from phase arbitrariness of data set which is critical for building MLPs for ESMD. As in the conventional ESMD simulations, the phase of NACVs can be chosen by comparing the sign of NACVs in the previous time step during ESMD simulations.
We note that the adiabatic state energy and gradient can be rewritten by $\Delta^2$ instead of $\Delta$ as 

\begin{equation}
    E_{\pm}^{SSR}= \frac{1}{2}((E^{PPS}+E^{OSS}){\pm}\sqrt{(E^{PPS}-E^{OSS})^2+4\Delta^2})
\end{equation}

and

\begin{eqnarray}
    \nabla E_{-}^{SSR} &= & n^2((E_{-}^{SSR}-E^{OSS})^2\nabla E^{PPS}+\Delta^2 \nabla E^{OSS}+(E_{-}^{SSR}-E^{OSS})\nabla \Delta^2)
   \nonumber \\
    \nabla E_{+}^{SSR} &= & m^2((E_{+}^{SSR}-E^{OSS})^2\nabla E^{PPS}+\Delta^2 \nabla E^{OSS}+(E_{+}^{SSR}-E^{OSS})\nabla \Delta^2).
\end{eqnarray}

In this study, we employed the SSR(2,2) approach with two electronic states only as an application of our machine learning scheme since currently only SSR(2,2) can offer analytic gradients for NAMD simulations. A two-state Hamiltonian effectively describes various photochemical reactions involving nonadiabatic coupling between the S$_0$ and S$_1$ states\cite{filatov_cychexadiene_2018, psb3dyn_2018, filatov_motor_2019, filatov_motor_2024}, but it can be extended to a three-state Hamiltonian through the training of phaseless functions (see Supporting Information S2 for details).

\subsection{Machine learning potential for phaseless quantities}
We employed the equivariant message-passing MLP, NequIP\cite{batznerEquivariantGraphNeural_2022}, implemented in the \texttt{Bayesian Atoms Modeling (BAM)} package, to train three key properties: $E^{PPS}$, $E^{OSS}$, and $\Delta^2$. Although a single MLP could theoretically model all properties simultaneously, task interference, where gradients from different tasks conflict, usually hinders learning\cite{liu2021}. And for this reason, we trained three independent MLPs for these diabatic properties, assuming that each property is best represented as a bijective mapping between molecular geometries and output. In training process, the analytical gradients of $\Delta^2$, denoted as $\nabla\Delta^2$, were constructed as $\nabla\Delta^2=2\Delta\nabla\Delta$ from $\Delta$ and $\nabla\Delta$ in the data set, which is also phaseless. To ensure that $\Delta^2$ remained positive, we applied inductive bias layers, such as \texttt{ReLU} activation function, enforcing a non-negative value constraint.
During the MLP prediction for $\Delta^2$, a small shift correction of $5 \times 10^{-5}$ eV$^2$ was applied which arose primarily because the ReLU activation function constrains $\Delta^2$ predictions to non-negative values (including zero), making it numerically unfeasible to predict exactly zero.



A database was generated consisting of 48,750 data points (40,000, 4,375, and 4,375 geometries for training, validation, and test, respectively), randomly selected from 50 trajectories obtained using exact factorization-based surface hopping dynamics (SHXF)\cite{jkha_deco_2018}. These trajectories were computed using the SSR(2,2) formalism combined with the $\omega$PBEh\cite{rohrdanz_2009}/6-31G*\cite{krishnan_1980} level of theory, with a time step of 0.24 fs over a total simulation duration of 300 fs.
Models were trained on a single NVIDIA H100 GPU, with the number of training epochs of 10,000. Averaged training time for three MLPs is 278 GPU-hours.
Each model was built with 5 interaction layers and 64 uncoupled feature channels, with $l_{\text{max}} = 2$ (Supporting Information S3 and S4).
In \texttt{e3nn} notation,\cite{e3nn_paper} this is represented as ``64x0o + 64x0e + 64x1o + 64x1e + 64x2o + 64x2e''. 
For all models, radial features were generated using 8 Bessel basis functions and a polynomial envelope with $p=2$ to handle the interatomic interaction within the cutoff.\cite{gasteigerDirectionalMessagePassing2022} 
These radial features are fed to a multi-layer perceptron \( \mathcal{M} \) with layer sizes [64, 64, 64, 1152], using \texttt{SiLU} activation functions in the hidden layers.  After the interaction layers, node energies or coupling elements were predicted using a single-layer \( \mathcal{M} \) with 16 hidden dimensions. 
A 6 \AA~ cutoff was applied to all molecules.  
The loss function used in training is defined as: 
\begin{equation}
\mathcal{L} = \frac{\lambda_E}{B} \sum_{b=1}^{B} \left( \hat{E}_b - E_b \right)^2 + \frac{\lambda_F}{3BN} \sum_{i=1}^{B \cdot N} \sum_{\alpha=1}^{3} \left( -\frac{\partial \hat{E}}{\partial r_{i,\alpha}} - F_{i,\alpha} \right)^2,
\end{equation}
where $B$ denotes the number of batches, $N$ the number of atoms in the batch, $E_b$ the target value ($E^{PPS}$, $E^{OSS}$, and $\Delta^2$), $\hat{E}_b$ the predicted value, and $F_{i,\alpha}$ the target minus gradient on atom $i$ in the direction $\alpha \in \{\hat{x}, \hat{y}, \hat{z}\}$. 
The weights $\lambda_E$ and $\lambda_F$ were optimized to 1 and 100 (for $\Delta^2$, 50), respectively.

The models were trained using the AMSGrad variant\cite{reddiConvergenceAdam2019} of the Adam optimizer,\cite{kingmaAdamMethodStochastic2017} with the default parameters of $\beta_1 = 0.9$, $\beta_2 = 0.999$, and $\epsilon = 10^{-8}$ (Supporting Information S5). We set the initial learning rate to 0.01 and used a batch size of 5. 
To adjust the learning rate, we applied an on-plateau scheduler based on the validation loss, with a patience of 500 epochs and a decay factor of 0.9. 
Additionally, we used an exponential moving average with a weight of 0.99 for evaluation on the validation set and for the final model.

\begin{figure}[!ht]
    \centering
    \includegraphics[scale=0.28]{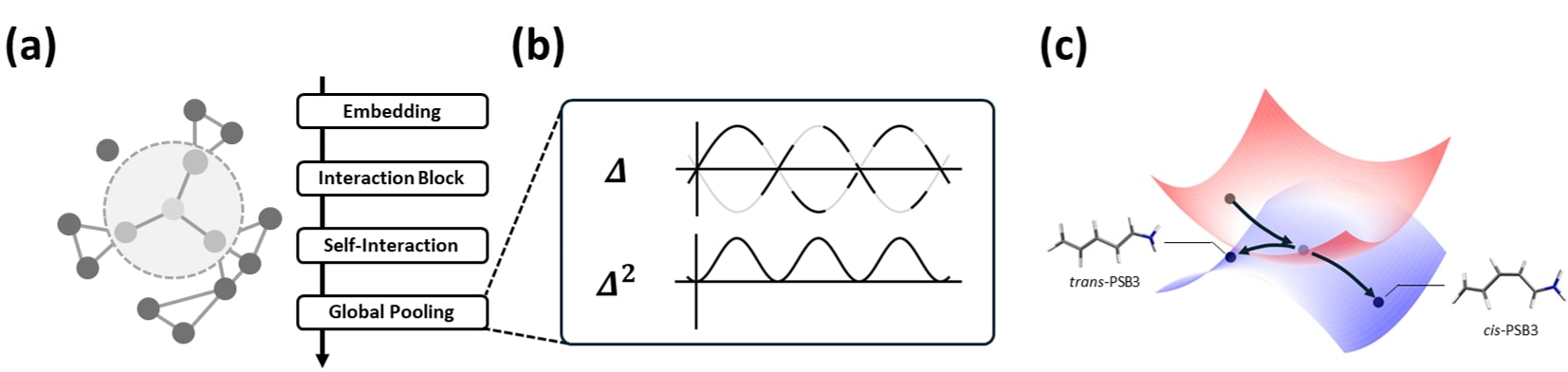}
    \caption{Overall schematic for this work. (a) Architecture of the NequIP-BAM model. (b) Global pooling block: \texttt{ReLU} activation enables learning of the phaseless coupling term $\Delta^2$(bottom), while the identity function shows discontinuities (top). (c) Nonadiabatic molecular dynamics schematic}
    \label{fig:figure1}
\end{figure}

The NequIP-BAM model architecture consists of four major blocks: embedding block, interaction block, self-interaction block, and global pooling block (Figure 1(a)). The embedding block maps atomic chemical properties and positional information into vector space, enabling efficient processing by the model. The interaction block models inter-atomic interactions while maintaining rotational equivariance. The self-interaction block updates features by applying identical weights for each atom and uses different weights according to rotational orders to maintain equivariance. Finally, the global pooling block aggregates features from each atom to generate final predictions. In particular, the global pooling block effectively learns the phaseless coupling term $\Delta^2$ by applying \texttt{ReLU} activation function, ensuring positive values (Figure 1(b)). The trained model enables nonadiabatic molecular dynamics that can clearly distinguish between $trans$-PSB3 and $cis$-PSB3 isomers, accurately predicting their positions and dynamical behavior on the potential energy surfaces (Figure 1(c)).

\section{Results and discussion}\label{sec:result}
\subsection{Model evaluation}

We first evaluated the models for their ability to predict diabatic properties, specifically $E^{PPS}$, $E^{OSS}$, and $\Delta^2$. 
As shown in Table 1 and Figure 2, the MLP predictions for diabatic properties are highly accurate. The MAE for the energy predictions of $E^{PPS}$ and $E^{OSS}$ is approximately 0.040 and 0.058 kcal/mol, respectively, while the gradient errors are about 0.063 and 0.080 kcal/mol/Å. Both $E^{PPS}$ and $E^{OSS}$ show high accuracy, surpassing the threshold for chemical accuracy. However, the error for $E^{OSS}$ is slightly larger than that of $E^{PPS}$, as shown in Table 1 and Figure S1, likely due to the greater difficulty in mapping excited-state PES from molecular geometry compared to ground-state PES. 

\begin{table}[!ht]
    \centering
    \begin{tabular}{c|ccc|ccc}
        \hline 
         && Energy &&& Gradient & \\
         & MAE & RMSE & R$^2$ & MAE & RMSE & R$^2$ \\
        \hline 
                $E^{PPS}$ & 0.040 & 0.072 & 0.999 & 0.063 & 0.181 & 0.999 \\
                $E^{OSS}$ & 0.058 & 0.111 & 0.999 & 0.080 & 0.259 & 0.999 \\
                $\Delta^2$& 0.463 & 0.810 & 0.999 & 1.135 & 2.626 & 0.998 \\
        \hline 
    \end{tabular}
    \caption{Mean Absolute Error (MAE),  Root-Mean-Square Error (RMSE), and R$^2$ values of the NequIP-BAM on the test dataset. $E^{PPS}$ and $E^{OSS}$ are given in kcal/mol, and $\Delta^2$ values are in (kcal/mol)$^2$.}
    \label{tab:table_x}
\end{table}

\begin{figure}[!ht]
    \centering
    \includegraphics[scale=0.5]{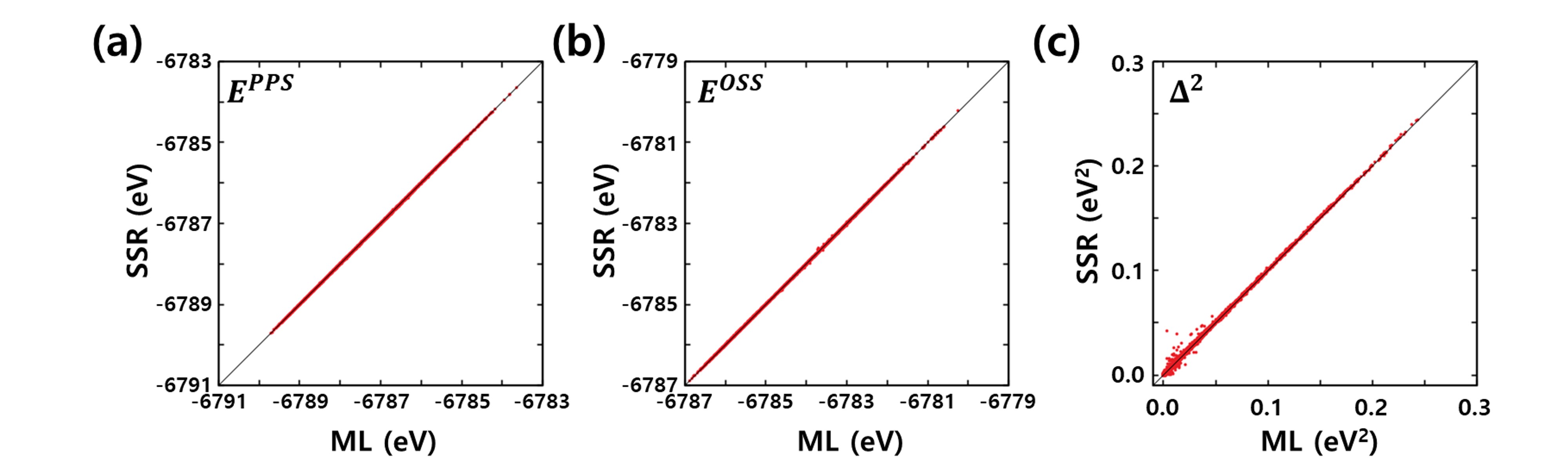}
    \caption{Comparison of diabatic Hamiltonian elements (in eV) between predictions from the NequIP-BAM model and reference SSR(2,2) values: (a) $E^{PPS}$ (b) $E^{OSS}$ (c) $\Delta^2$.}
    \label{fig:figure2}
\end{figure}

Increasing the number of layers in the Interaction Block could potentially enhance accuracy. 
Indeed, when we tested the model's accuracy with varying numbers of layers, we observed nearly a twofold improvement in accuracy with the addition of an extra layer for $E^{OSS}$. We also compared the model with different numbers of feature channels, and the results showed that increasing the number of features was less effective compared to increasing the number of layers. However, due to computational cost constraints, we used 5 layers and 64 feature channels in our final model. Table S1 presents the results for different layer configurations, while Table S2 provides the results for various feature channels. For $\Delta^2$, the MAE of the energy is approximately 0.463 (kcal/mol)$^2$ for the NequIP-BAM model.
In addition, we observe a larger error for the small $\Delta^2$ region (Figure S1(c)). This occurs because $\Delta^2$ becomes small near CIs as well as in the region far away from CIs with a large energy gap $|E_{PPS}-E_{OSS}|$. Thus, vastly different molecular geometries can result in similar $\Delta^2 \approx 0$ values, making it inherently more challenging for the MLP to distinguish between these geometries. Our approach is notable not only for its accuracy but also for successfully resolving the scattered pattern observed in previous work\cite{ESNAMD_2021}, which used a phaseless loss function. This result demonstrates that squaring the double-valued $\Delta$ is an effective strategy for generating a continuous, smooth function that can be efficiently trained. This suggests that the model is no longer affected by the double-valued phase problem and can predict nonadiabatic properties with a similar level of accuracy to conventional models.


Next, we compared the effects of different activation functions applied to the bias layer.
During $\Delta^2$ training, we observed negative output values from MLP, which are undesirable given the positive nature of the $\Delta^2$.
Therefore, \texttt{ReLU} and \texttt{SiLU} activation functions were applied during the $\Delta^2$ training and these results are presented in Table 2.
The \texttt{ReLU} activation function improves the accuracy compared against the model without the activation function.
The MAE of $\Delta^2$ with \texttt{ReLU} activation is 0.463 (kcal/mol)$^2$, compared to 0.634 (kcal/mol)$^2$ for the model with the identity activation.
When the reference value was near zero value, using the identity activation function occasionally resulted in negative predictions for $\Delta^2$, which are theoretically impossible. In contrast, the \texttt{ReLU} activation function produced positive predictions due to its bias. This suggests that \texttt{ReLU} is effective in reducing prediction errors for $\Delta^2$.
While \texttt{ReLU} showed a smaller test error compared to the identity function, the \texttt{SiLU} activation resulted in a significantly larger error of 29.090 (kcal/mol)$^2$. The issue likely comes from \texttt{SiLU}’s handling of negative values. It is designed to prevent the dying \texttt{ReLU} problem, where neurons become inactive and gradients vanish\cite{maasrectifier}. However, this conflicts with the strict requirement for $\Delta^2$ to remain positive.
Based on these findings, we concluded that the \texttt{ReLU} activation function, combined with the NequIP-BAM model, is the most suitable MLP architecture for training PSB3 model. This architecture was applied in all subsequent training and ML-ESMD simulations.

\begin{table}[!ht]
    \centering
    \begin{tabular}{c|ccc|ccc}
        \hline 
         && Energy &&& Gradient & \\
         & MAE & RMSE & R$^2$ &MAE & RMSE & R$^2$ \\
        \hline 
                Identity  & 0.634 & 1.084 & 0.998 & 1.124 & 2.186 & 0.999 \\
                \texttt{ReLU} & 0.463 & 0.810 & 0.999 & 1.135 & 2.626 & 0.998 \\
                \texttt{SiLU} & 29.090 & 36.472 & 0.116 & 2.723 & 4.961 & 0.994 \\
        \hline 
    \end{tabular}
    \caption{Mean Absolute Error (MAE),  Root-Mean-Square Error (RMSE), and R$^2$ values of the NequIP-BAM for $\Delta^2$ prediction on the test dataset. The units are expressed in (kcal/mol)$^2$.}
    \label{tab:table_2}
\end{table}

While overall model performance is important, accurately predicting the minimum-energy conical intersection (MECI) structure is critical due to its central role in nonadiabatic dynamics involving multiple electronic states. The degeneracy of the two adiabatic surfaces at the CI is lifted by two branching plane vectors: the difference gradient vector $\vec{g}$ and the coupling derivative vector $\vec{h}$. To assess the model’s accuracy at the CI, we compared the predicted branching plane vectors and MECI structures with their respective references. Figure 3 shows the MECI structures and two branching plane vectors for the CI structure, as predicted by the NequIP-BAM model and calculated by SSR(2,2)/$\omega$PBEh/6-31G* (See Supporting Information S9 for the geometry of CI and two branching plane vectors). As shown in Figure 3(a), the MECI structures predicted by the NequIP-BAM model and calculated using SSR(2,2) are nearly indistinguishable. The $\vec{g}$-vector represents the bond elongation motion of the terminal C=C bond, while the $\vec{h}$-vector corresponds to the torsional motion of the central C=C bond. These two motions are known to significantly contribute to molecular dynamics at the CI in PSB3, as previously reported\cite{psb3dyn_2018}. The predicted vectors at the CI region match the direction and magnitude of the reference values. 
In addition, we can also predict the adiabatic potential energy surface and discontinuity of real-valued eigenvectors around the CI in the branching plane (Figure S2), which is critical to assess Berry phase due to conical topology of CIs.

\begin{figure}[!ht]
    \centering
    \includegraphics[width=6in]{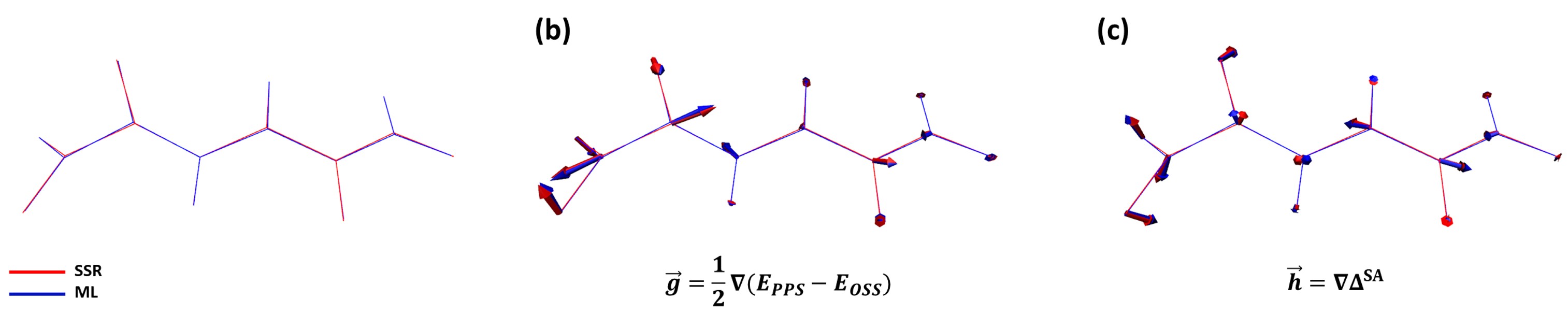}
    \caption{MECI structure and its branching plane vectors: (a) Comparison of MECI structures predicted by the NequIP-BAM model (blue) and calculated by SSR(2,2)/$\omega$PBEh/6-31G* (red). (b) difference gradient vector ($\Vec{g}$) and (a) coupling derivative vector ($\Vec{h}$). The red and blue arrows denote vectors calculated by SSR(2,2)/$\omega$PBEh/6-31G* and predicted by the NequIP-BAM model respectively.
    }
    \label{fig:figure3}
\end{figure}

Since NACV plays a key role in nonadiabatic dynamics, we evaluated the prediction accuracy of NACV in the vicinity of the CI region by comparing the NequIP-BAM model with the reference SSR(2,2) values.
NACVs were calculated using eq \ref{eq:nacv}, with all components computed from the MLP predictions of $E^{PPS}$, $E^{OSS}$, $\Delta^2$, and their respective gradients. We divided the NACV test set into two groups: one near the CI region (404 geometries) and one away from it (2,285 geometries)(Figure 4). The NACV vector contains 42 components. Thus, Figure 4 includes 16968 elements in the vicinity of the CI and 95970 elements distant from the CI. The division was based on an adiabatic energy difference of 0.5 eV. The predicted vector elements from the model closely matched their reference values, with small errors for most points (Figure S1(d)). Accurate NACV prediction near the CI  is crucial for describing correct state transitions. Additionally, accurate prediction of NACVs far from the CI region is also important to prevent erroneous transitions in regions where the coupling between electronic states is weak. We confirm that the NACV predictions are not overestimated and remain accurate when it's away from the CI (Figure 4).
 In surface hopping dynamics, where hopping probabilities are directly influenced by NACVs, the precise prediction of NACVs by our MLP strengthens the reliability of state transitions in our ML-ESMD simulations.

\begin{figure}[!ht]
    \centering
    \includegraphics[scale=0.8]{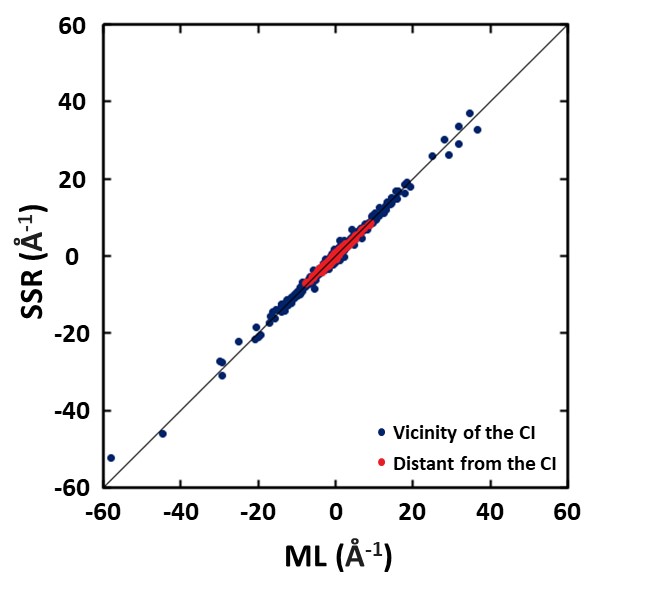}
    \caption{Scatter plots of NACV predictions from the NequIP-BAM model compared to the reference SSR(2,2) values: Adiabatic energy difference < 0.5 eV (blue) and adiabatic energy differences $\geq$ 0.5 eV (red).}
    \label{fig:figure4}
\end{figure}

\subsection{Nonadiabatic molecular dynamics}

After successfully demonstrating that our MLP model accurately reproduces static properties, we ran nonadiabatic dynamics simulations. We used MLP for nonadiabatic MD simulations based on the SHXF method\cite{jkha_deco_2018}, implemented in the PyUNIxMD package\cite{pyunixmd_2021, pyunixmd_2022}.
For the initial conditions, 100 nuclear configurations and their velocities were newly sampled from the optimized ground-state geometry of the $trans$-PSB3 molecule, randomly selected from Wigner distribution on 300K. 
Dynamics were performed over a total of 300 fs with a timestep of 0.24 fs.
The nonadiabatic relaxation of PSB3 from the $S_1$ state to the ground state via MECI involves changes in the torsion angle of the central C=C bond, which leads to the differentiation to the $cis$ and $trans$ isomers\cite{psb3dyn_2018}.

We tracked the dihedral angle of the central C=C bond (Figure 5). The dihedral angle trajectories clearly show the separation into the $cis$ and $trans$ isomers. Although the results from 100 trajectories may not be statistically significant, the final cis-to-trans ratio of 61:39 is close to the reference ratio of 63:37\cite{psb3dyn_2018}. Current work improves on the previous work\cite{ESNAMD_2021}, which reported a ratio of 58.5:41.5. Additionally, the comparison of ML-based and SSR-based trajectories reveals that the molecular motions and final yields are in strong agreement between the two approaches (Figure 4 and 5).

We also confirm that the average electronic population evolutions from ML and reference SSR dynamics simulations are in agreement (Figure 5(b)).
As described in Figure S3, the magnitude of nonadiabatic couplings from ML prediction increases properly as the two adiabatic energies close to each other.
Averaged populations $P_i$ and $p_i$ are calculated from the running state and BO population, where $P_i=N_i/N_{traj}$ and $p_i=\sum\limits_{I}^{N_{traj}}\rho_{ii}(t)/N_{traj}$, respectively.
Here, $N_i$, $N_{traj}$, and $\rho_{ii}$ represent the number of trajectories running on the $i$-th electronic state, the total number of trajectories, and a BO population at time $t$, respectively.
The agreement between the two populations highlights the MLP’s ability to accurately capture electron-nuclear correlations. Furthermore, the population trends observed in our SHXF/ML dynamics closely match those from the SHXF/SSR(2,2) simulations, despite the ML model being approximately 100 times faster. 

\begin{figure}[!ht]
    \centering
    \includegraphics[scale=0.45]{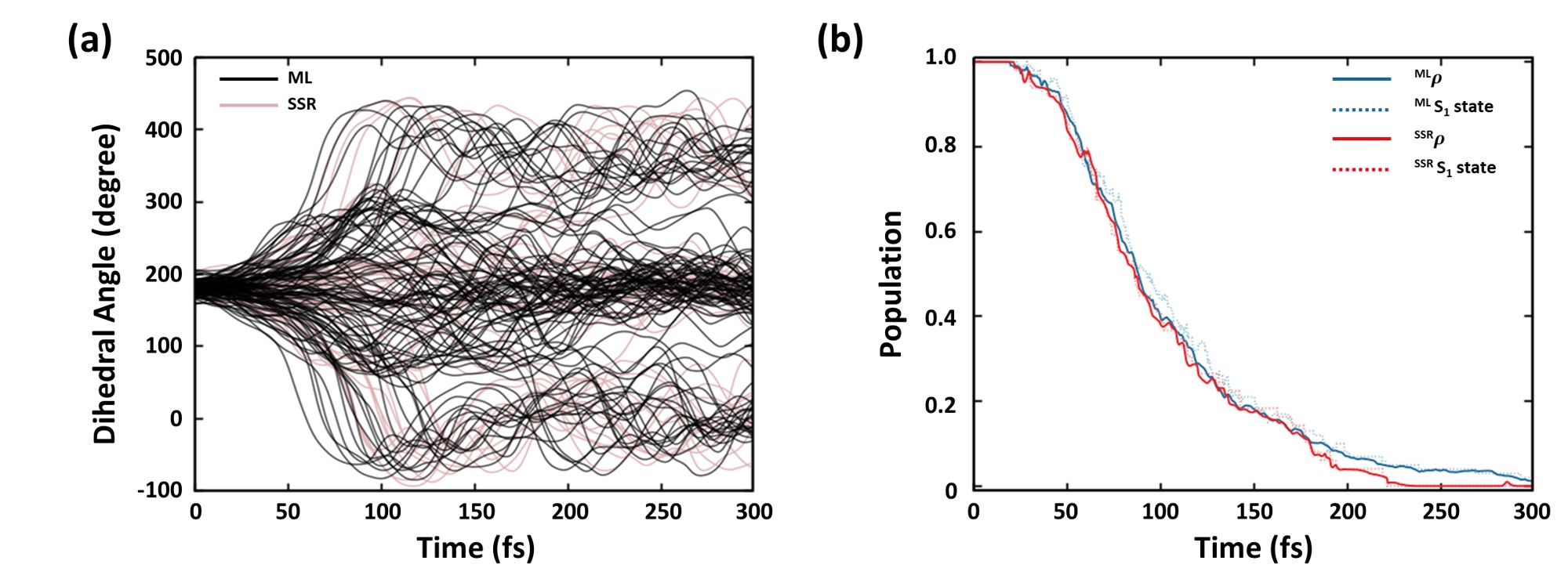}
    \caption{Analysis of the dynamics of photoisomerization of PSB3: (a) Dihedral angle of the central C=C bond of PSB3 over time in individual trajectories. Black trajectories represent ML-based results, while red trajectories represent reference SSR-based dynamics. (b) Average electronic population with the corresponding time. The blue and red line represents ML and reference SSR population evolution, respectively. The solid line represents the BO population and the dashed line represents the averaged running state.}
    \label{fig:figure5}
\end{figure}

\section{Conclusion}\label{sec:conc}

We propose a phaseless ML-ESMD approach by building MLPs for the diabatic Hamiltonian elements—$E^{PPS}$, $E^{OSS}$, and $\Delta^2$—obtained from the SSR (2,2) method. This formalism addresses the phase arbitrariness in ML-ESMD. By focusing on predicting diabatic Hamiltonian elements rather than directly on the NACV, we avoid the singularities at CIs. Additionally, we reformulate the NACV and its gradients by replacing the phase-dependent off-diagonal elements with their squared values $\Delta^2$. MLP models trained with this approach show high accuracy without any discontinuities or divergence. We show that the ML-ESMD of the PSB3 system agrees well with the reference ab initio simulations. 
We expect that our ML approach, when combined with state-of-the-art MLPs, will aid in the development of multi-state universal MLPs, applicable to systems encompassing elements across the entire periodic table, which remains as future work.\cite{batatia2024foundationmodelatomisticmaterials} These models could be integrated into de novo molecular design, allowing for the exploration of ever more complex and challenging problems in the field.

\section{Code availability}
The input and output data files associated with this study and all analysis can be found on GitHub at \url{https://github.com/myung-group/Data_phaseless_namd} and Zenodo\cite{psb3_data}. The source code for the Bayesian Atoms Modeling (BAM) package is available on GitHub at \url{https://github.com/myung-group/BAM-Public}. 

\begin{acknowledgement}
This research was supported by the National Research Foundation of Korea (NRF) funded by the Korean government (Ministry of Science and ICT(MSIT))(NRF-2023M3K5A1094813, RS-2023-00257666, RS-2024-00455131, NRF-2022R1C1C1010605) and by Institute for Basic Science (IBS-R036-D1). SYW and CWM acknowledge the support from Brain Pool program funded by the Ministry of Science and ICT through the National Research Foundation of Korea (No. RS-2024-00407680). CWM, THP, and SYW are grateful for the computational support from the Korea Institute of Science and Technology Information (KISTI) for the Nurion cluster (KSC-2023-CRE-0502, KSC-2023-CRE-0355, KSC-2023-CRE-0059, KSC-2022-CRE-0424, KSC-2022-CRE-0115, KSC-2022-CRE-0113, KSC-2021-CRE-0542, KSC-2021-CRE-0496, KSC-2022-CRE-0429, KSC-2022-CRE-0469, KSC-2023-CRE-0050, KSC-2023-CRE-0251, KSC-2023-CRE-0261, KSC-2023-CRE-0332, KSC-2023-CRE-0472, KSC-2023-CRE-0501, KSC-2024-CRE-0358). Computational work for this research was partially performed on the Olaf supercomputer supported by IBS Research Solution Center.
\end{acknowledgement}
\begin{suppinfo}
See the supplementary material for a detailed compilation of the obtained results as well as further data and analysis to support the points made throughout the text. 
\end{suppinfo}

\bibliography{achemso}

\end{document}


\clearpage

\tableofcontents{}

\clearpage

\newpage
\subsection*{S1. Derivation of Equations (8) and (10)}
In this section, we provide detailed derivation of Equation (8) in the main text. From the SSR(2,2) formalism
\begin{equation}
    \begin{pmatrix}
    E_{PPS} & \Delta \\
    \Delta & E_{OSS} \\
    \end{pmatrix}
    \begin{pmatrix}
    a_{00} & a_{01} \\
    a_{10} & a_{11} \\
    \end{pmatrix}
    =
    \begin{pmatrix}
    E_{-}^{SSR} & 0 \\
    0 & E_{+}^{SSR} \\
    \end{pmatrix}
    \begin{pmatrix}
    a_{00} & a_{01} \\
    a_{10} & a_{11} \\
    \end{pmatrix}.
\end{equation}
The eigenvectors corresponding to eigenvalues $E^{SSR}_+$ and $E^{SSR}_-$ with $E^{SSR}_- < E^{SSR}_+$ can be written as
\begin{equation}\label{S2}
    \begin{pmatrix}
    a_{00} \\
    a_{10} \\
    \end{pmatrix}
    =
    n
    \begin{pmatrix}
    E^{SSR}_- - E_{OSS} \\
    \Delta  \\    
    \end{pmatrix}
\end{equation}
and
\begin{equation}\label{S3}
    \begin{pmatrix}
    a_{01} \\
    a_{11} \\
    \end{pmatrix}
    =
    m
    \begin{pmatrix}
    E^{SSR}_+ - E_{OSS} \\
    \Delta  \\    
    \end{pmatrix}
\end{equation}
respectively, where $n$ and $m$ are chosen as real-valued normalization constants, 

\noindent $n=\pm1/\sqrt{(E^{SSR}_- - E_{OSS})^2 + \Delta^2}$ and  $m=\pm1/\sqrt{(E^{SSR}_+ - E_{OSS})^2 + \Delta^2}$, respectively.
Thus, the real-valued eigenvectors can be chosen as positive or negative, given real-valued $\Delta$.
By putting the above equations in Equation (7) in the main text, we can obtain
\begin{align}
    d_{01} &= \frac{nm}{E_+ - E_-}((-\Delta^2)\nabla(E_{PPS}-E_{OSS}) +(E_{PPS} - E_{OSS})\Delta\nabla\Delta) \nonumber\\
           &= \frac{nm}{E_+ - E_-}((-\Delta^2)\nabla(E_{PPS}-E_{OSS}) +\frac{1}{2}(E_{PPS} - E_{OSS})\nabla\Delta^2)
\end{align}               
since $2\Delta\nabla\Delta$ can be expressed as $\nabla\Delta^2$.
Thus, once we construct machine learning models for phaseless quantities $E_{PPS}$, $E_{OSS}$, and $\Delta^2$, we can obtain adiabatic energies and NACVs directly while the phase of NACVs can be matched with NACVs from previous time steps during the conventional excited state molecular dynamics simulations.
If we try to construct the continuous diabatic Hamiltonian, we can choose the sign of $\Delta$ based on the gradient of $\Delta$ to make continuous changes in $\nabla\Delta =\nabla\Delta^2/2\Delta$.

For gradients of adiabatic state energies $E^{SSR}_\pm$, we can put the eigenvectors in equations (\ref{S2}) and (\ref{S3}) in Hellmann-Feynman gradients
\begin{align}
    \nabla E^{SSR}_- &= a^2_{00}\nabla E_{PPS} + a^2_{10}\nabla E_{OSS} + 2a_{00}a_{10}\nabla\Delta \\
    \nabla E^{SSR}_+ &= a^2_{01}\nabla E_{PPS} + a^2_{11}\nabla E_{OSS} + 2a_{01}a_{11}\nabla\Delta
\end{align}
to obtain
\begin{align}
    \nabla E_{-}^{SSR} &= & n^2((E_{-}^{SSR}-E^{OSS})^2\nabla E^{PPS}+\Delta^2 \nabla E^{OSS}+(E_{-}^{SSR}-E^{OSS})\nabla \Delta^2)
   \nonumber \\
    \nabla E_{+}^{SSR} &= & m^2((E_{+}^{SSR}-E^{OSS})^2\nabla E^{PPS}+\Delta^2 \nabla E^{OSS}+(E_{+}^{SSR}-E^{OSS})\nabla \Delta^2).
\end{align}

\clearpage
\subsection*{S2. Extension to three-state Hamiltonians}
In this section, we explain the extension of our machine learning strategy to three-state Hamiltonians. 
If we focus on real-valued Hamiltonians, the diabatic Hamiltonian $H$ becomes a real symmetric matrix
\begin{align}
H = 
    \begin{pmatrix}
    E_{PPS} & \Delta_{01} & \Delta_{02}\\
    \Delta_{01} & E_{OSS} & \Delta_{12} \\
    \Delta_{02} & \Delta_{12} & E_{DES} \\    
    \end{pmatrix}
    =
\begin{pmatrix}
    a & b & c \\ b & d & e \\ c & e & f
\end{pmatrix}
\end{align}
where we put the Hamiltonian for SSR(3,2) approach as an example. The SSR(3,2) approach deals with a doubly excited singlet (DES) state in addition to SSR(2,2)\cite{REKS32_2021}.

The eigenvalues $\lambda_i$ can be written as
\begin{align}
\lambda_1 &= \frac{{(D + \sqrt{4 A^3 + B^2} + C)}^{1/3}}{3\cdot 2^{1/3}} - \frac{2^{1/3} A}{3 (D + \sqrt{4 A^3 + B^2} + C)^{1/3}} + \Lambda\nonumber\\
\lambda_2 &= -\frac{(1 - i \sqrt{3}){(D + \sqrt{4 A^3 + B^2} + C)}^{1/3}}{6\cdot 2^{1/3}} + \frac{(1 + i \sqrt{3}) A}{3\cdot 2^{2/3} (D + \sqrt{4 A^3 + B^2} + C)^{1/3}} + \Lambda\nonumber\\
\lambda_3 &= -\frac{(1 + i \sqrt{3}) {(D + \sqrt{4 A^3 + B^2} + C)}^{1/3}}{6\cdot 2^{1/3}} + \frac{(1 - i \sqrt{3}) A}{3\cdot 2^{2/3} (D + \sqrt{4 A^3 + B^2} + C)^{1/3}} + \Lambda
\end{align}
where
\begin{align}
 A =& -a^2 + a d + a f - 3 b^2 - 3 c^2 - d^2 + d f - 3 e^2 - f^2\nonumber\\
B =& 2 a^3 - 3 a^2 d - 3 a^2 f + 9 a b^2 + 9 a c^2 - 3 a d^2 + 12 a d f - 18 a e^2 - 3 a f^2 + 9 b^2 d - 18 b^2 f \nonumber\\
&+ 54 b c e - 18 c^2 d + 9 c^2 f + 2 d^3 - 3 d^2 f + 9 d e^2 - 3 d f^2 + 9 e^2 f + 2 f^3\nonumber\\
C =& 9 a b^2 + 9 a c^2 - 3 a d^2 + 12 a d f - 18 a e^2 - 3 a f^2 + 9 b^2 d - 18 b^2 f + 54 b c e \nonumber\\
&- 18 c^2 d + 9 c^2 f + 2 d^3 - 3 d^2 f + 9 d e^2 - 3 d f^2 + 9 e^2 f + 2 f^3\nonumber\\
D =& 2 a^3 - 3 a^2 d - 3 a^2 f \nonumber\\
\Lambda =& \frac{1}{3}(a+d+f).
\end{align}
Here, complex values appear in general since $4A^3+B^2$ can be negative, but the eigenvalues become real as a result. 
Importantly, we note that the eigenvalues depend only on phaseless quantities $a$, $d$, $f$, $b^2$, $c^2$, $e^2$, and $bce$ where $bce$ can be proven as invariant with respect to the sign change of basis states.
Still, the function $bce$ behaves as a normal function (not positive definite) which can change the sign depending on the nuclear configuration. While training $bce$, an activation function may be not necessary.
Therefore, once we train diagonal elements ($a$, $d$, and $f$) and the square of off-diagonal elements ($b^2$, $c^2$, and $e^2$) in addition to the product of three off-diagonal elements, $bce$, we can obtain invariant eigenvalues, $E_0 < E_1 < E_2$, and corresponding eigenvectors, $|i\rangle$ with $i=0,1,2$, at a given nuclear configuration by diagonalizing the Hamiltonian $H^{ML}$
\begin{align}
H^{ML} =
\begin{pmatrix}
    \tilde{a} & \tilde{b} & \tilde{c} \\
    \tilde{b} & \tilde{d} & \tilde{e} \\
    \tilde{c} & \tilde{e} & \tilde{f}
\end{pmatrix}
\end{align}
where $\tilde{a} = a$, $\tilde{d} = d$, and $\tilde{f} = f$, $\tilde{b} = \sqrt{b^2}$, $\tilde{c} = \sqrt{c^2}$, and $\tilde{e} = \text{sgn(}{bce}\text{)}\sqrt{e^2}$.
The eigenvectors have phase freedom naturally as in a 2x2 Hamiltonian.
Once we obtain eigenvalues and eigenvectors, we can calculate NACVs $d_{ij,\nu}$ as
\begin{align}
    d_{ij} = \frac{\langle i|\nabla\nu H|j\rangle}{\lambda_j-\lambda_i}
\end{align}
where the gradient of off-diagonal elements can be calculated by $\nabla\Delta = \nabla\Delta^2/2\Delta$.
The sign of NACVs can be chosen during excited state molecular dynamics simulations as explained in the previous section.

For example, $d_{01}$ can be written as
\begin{align}
    d_{01} &= \frac{1}{E_1^{SSR} - E_0^{SSR}}((a_{00}a_{01}\nabla E_{PPS} + a_{10}a_{11}\nabla E_{OSS}+a_{20}a_{21}\nabla E_{DES}) \nonumber\\
    &+ (a_{00}a_{11} + a_{01}a_{10})\nabla\Delta_{01}+ (a_{00}a_{21} + a_{01}a_{20})\nabla\Delta_{02}+ (a_{10}a_{21} + a_{20}a_{11})\nabla\Delta_{12})    
\end{align}
where the coefficients for adiabatic states $|0\rangle$ and $|1\rangle$ can be written as $(a_{00},a_{10},a_{20}) = n(E_{D-0}(\Delta_{01}\Delta_{12}-\Delta_{02}E_{O-0}),E_{D-0}(\Delta_{01}\Delta_{02}-\Delta_{12}E_{P-0}),\Delta_{02}^2E_{O-0}+\Delta_{12}^2E_{P-0}-2\Delta_{01}\Delta_{02}\Delta_{12})$ and $(a_{01},a_{11},a_{21}) = m(E_{D-1}(\Delta_{01}\Delta_{12}-\Delta_{02}E_{O-1}),E_{D-1}(\Delta_{01}\Delta_{02}-\Delta_{12}E_{P-1}),\Delta_{02}^2E_{O-1}+\Delta_{12}^2E_{P-1}-2\Delta_{01}\Delta_{02}\Delta_{12})$, respectively, with real-valued normalization constants $n$ and $m$ which can be positive or negative. Here, $E_{P/O/D-0/1}$ represents the energy difference between diabatic states ($P$, $O$, and $D$ represent PPS, OSS, and DES, respectively) and adiabatic states ($0$ and $1$), e.g. $E_{D-0} = E_{DES}-E_0$.

Then, the $d_{01}$ can be explicitly written as
\begin{align}
    d_{01} &= \frac{nm}{E_1^{SSR} - E_0^{SSR}}[[E_{D-0}E_{D-1}(\Delta_{01}^2\Delta_{12}^2-\Delta_{01}\Delta_{02}\Delta_{12}(E_{O-0}+E_{O-1})+\Delta_{02}^2E_{O-0}E_{O-1})]\nabla E_{PPS} \nonumber \\
            &\hspace{1em}+[E_{D-0}E_{D-1}(\Delta_{01}^2\Delta_{02}^2-\Delta_{01}\Delta_{02}\Delta_{12}(E_{P-0}+E_{P-1})+\Delta_{12}^2E_{P-0}E_{P-1})]\nabla E_{OSS} \nonumber \\
            &\hspace{1em}+[(\Delta_{02}^2E_{O-0}+\Delta_{12}^2E_{P-0}-2\Delta_{01}\Delta_{02}\Delta_{12})(\Delta_{02}^2E_{O-1}+\Delta_{12}^2E_{P-1}-2\Delta_{01}\Delta_{02}\Delta_{12})]\nabla E_{DES} \nonumber \\
            &\hspace{1em}+ \frac{1}{2}E_{D-0}E_{D-1}[2\Delta_{01}\Delta_{02}\Delta_{12}-\Delta_{02}^2(E_{O-0}+E_{O-1})-\Delta_{12}^2(E_{P-0}+E_{P-1})\nonumber\\
            &\hspace{8em}+\frac{\Delta_{01}\Delta_{02}\Delta_{12}}{\Delta_{01}^2}(E_{O-0}E_{P-1}+E_{O-1}E_{P-0})]\nabla \Delta_{01}^2 \nonumber\\
            &\hspace{1em}+\frac{1}{2}[\frac{\Delta_{01}\Delta_{02}\Delta_{12}^3}{\Delta_{02}^2}(E_{D-0}E_{P-1}+E_{D-1}E_{P-0})-\Delta_{12}^2(E_{P-1}E_{O-0}E_{D-0}+E_{P-0}E_{O-1}E_{D-1}) \nonumber \\
            &\hspace{4em}+\Delta_{01}\Delta_{02}\Delta_{12}(2E_{D-0}E_{O-0}+2E_{D-1}E_{O-1}+E_{D-0}E_{O-1}+E_{D-1}E_{O-0}) \nonumber \\
            &\hspace{4em}-(E_{D-0}+E_{D-1})(2\Delta_{01}^2\Delta_{12}^2+\Delta_{02}^2E_{O-0}E_{O-1})]\nabla\Delta_{02}^2 \nonumber \\
            &\hspace{1em}+\frac{1}{2}[\frac{\Delta_{01}\Delta_{02}^3\Delta_{12}}{\Delta_{12}^2}(E_{D-0}E_{O-1}+E_{D-1}E_{O-0})-\Delta_{02}^2(E_{P-0}E_{O-1}E_{D-0}+E_{P-1}E_{O-0}E_{D-1}) \nonumber \\
            &\hspace{4em}+\Delta_{01}\Delta_{02}\Delta_{12}(2E_{D-0}E_{P-0}+2E_{D-1}E_{P-1}+E_{D-0}E_{P-1}+E_{D-1}E_{P-0}) \nonumber \\
            &\hspace{4em}-(E_{D-0}+E_{D-1})(2\Delta_{01}^2\Delta_{02}^2+\Delta_{12}^2E_{P-0}E_{P-1})]\nabla\Delta_{12}^2].
\end{align}
We note that $d_{01}$ can be obtained from $E_{PPS}$, $E_{OSS}$, $E_{DES}$, $\Delta_{01}^2$, $\Delta_{02}^2$, $\Delta_{12}^2$, and $\Delta_{01}\Delta_{02}\Delta_{12}$ which are phaseless, and the phase of $nm$ can be determined during excited state molecular dynamics simulations. The development of machine learning potentials with the above formulations remain as a future work.

\newpage
\subsection*{S3. Hyperparameter test model for NequIP-BAM}

\begin{table}[h!]
    \centering
    \begin{tabular}{cc|cc|cc}
        \hline 
         && \multicolumn{2}{c|}{Energy} & \multicolumn{2}{c}{Gradient} \\
         && MAE & RMSE & MAE & RMSE\\
        \hline 
         layer 3 & $E^{PPS}$ & 0.119 & 0.173 & 0.257 & 0.423 \\
                 & $E^{OSS}$ & 0.178 & 0.251 & 0.370 & 0.604 \\
                 & $\Delta^2$& 1.660 & 2.221 & 3.820 & 7.056 \\
         layer 5 & $E^{PPS}$ & 0.046 & 0.085 & 0.093 & 0.232 \\
                 & $E^{OSS}$ & 0.054 & 0.101 & 0.134 & 0.326 \\
                 & $\Delta^2$& 0.619	& 1.040	& 1.444	& 4.379 \\               
        \hline 
    \end{tabular}
    \caption{Mean Absolute Error (MAE) and  Root-Mean-Square Error (RMSE) values of the trained $\Delta^2$ model with different numbers of layers. $E^{PPS}$ and $E^{OSS}$ are given in kcal/mol, and $\Delta^2$ values are in (kcal/mol)$^2$.}
    \label{tab:table_S1}
\end{table}

\begin{table}[h!]
    \centering
    \begin{tabular}{cc|cc|cc}
        \hline 
         && \multicolumn{2}{c|}{Energy} & \multicolumn{2}{c}{Gradient} \\
         && MAE & RMSE & MAE & RMSE\\
        \hline 
         32 features & $E^{PPS}$ & 0.119 & 0.173 & 0.257 & 0.423 \\
                     & $E^{OSS}$ & 0.178 & 0.251 & 0.370 & 0.604 \\
                     & $\Delta^2$& 1.660 & 2.221 & 3.820 & 7.056 \\
         64 featrures & $E^{PPS}$ & 0.116 & 0.167 & 0.241 & 0.404\\
                      & $E^{OSS}$ & 0.155 & 0.218 & 0.325 & 0.550 \\
                      & $\Delta^2$&1.521 & 2.119	& 3.670	& 6.838\\              
        \hline 
    \end{tabular}
    \caption{Mean Absolute Error (MAE) and  Root-Mean-Square Error (RMSE) values of the trained $\Delta^2$ model with different numbers of features. $E^{PPS}$ and $E^{OSS}$ are given in kcal/mol, and $\Delta^2$ values are in (kcal/mol)$^2$.}
    \label{tab:table_S2}
\end{table}

\setlength{\abovedisplayskip}{0pt}  
\setlength{\belowdisplayskip}{0pt}  

\newpage
\subsection*{S4. Model Architecture}

We need to calculate the convolution filter \( S_m^{\left(l\right)}\left({\vec{r}}_{ij}\right) \),  
\[
S_m^{\left(l\right)}\left({\vec{r}}_{ij}\right) = R\left(r_{ij}\right) Y_m^{\left(l\right)}\left({\hat{r}}_{ij}\right).
\]  
When the dimension of spherical harmonics \( Y_m^{\left(l\right)} \) is defined as \( 64 \times 0o + 64 \times 0e + 64 \times 1o + 64 \times 1e + 64 \times 2o + 64 \times 2e \), it is necessary for the radial basis dimension to match the spherical harmonics. Therefore, we perform the multi-layer perceptron \( \mathcal{M} \) transformation from \( (n_{\text{atom}}, n_{\text{basis}}) \) to \( (n_{\text{atom}}, n_{\text{hidden}}) \).  

The output size of \( n_{\text{hidden}} (= 1152) \) was determined based on the specified representation in e3nn notation\cite{batznerEquivariantGraphNeural_2022}, \( 64 \times 0o + 64 \times 0e + 64 \times 1o + 64 \times 1e + 64 \times 2o + 64 \times 2e \). This corresponds to:  
\begin{itemize}
    \item \( 0o \) and \( 0e \): Each contributes 1 dimension.
    \item \( 1o \) and \( 1e \): Each contributes 3 dimensions.
    \item \( 2o \) and \( 2e \): Each contributes 5 dimensions.
\end{itemize}

Thus, the total dimensionality is calculated as follows:
\[
2 \cdot 64 \cdot 1 + 2 \cdot 64 \cdot 3 + 2 \cdot 64 \cdot 5 = 1152
\]
Here, each term accounts for the multiplicative combination of features, parity (even or odd), and the degree \( l_{\text{max}} \) of the spherical harmonics.

The structure of the \( \mathcal{M} \) with layer sizes \([64, 64, 64]\) reflects this design, where:  
\begin{itemize}
    \item \( I \) represents the input features of size \( \left[n_{\text{atom}},\ n_{\text{basis}}\right] \), where \( n_{\text{basis}} (= 8) \) is the number of radial basis functions.
    \item The transformations are applied via weight matrices \( W_1 \) of shape \([n_{\text{basis}}, 64]\), \( W_2 \) of shape \([64, 64]\), \( W_3 \) of shape \([64, 64]\), \( W_4 \) of shape \([64, n_{\text{hidden}}]\).
    \item The output \( O \) is of size \( \left[n_{\text{atoms}},\ n_{\text{hidden}}\right] \), consistent with the dimensionality required to handle the interaction channels represented in the e3nn format.
\end{itemize}

\newpage
\subsection*{S5. Hyperparameter of optimizer}
The AMSGrad algorithm is a variant of the Adam optimizer designed to improve convergence behavior by ensuring that the second moment estimate does not decrease during training. In AMSGrad, the first moment estimate \( m_t \) is updated as 
\[
m_t = \beta_1 m_{t-1} + \left(1 - \beta_1\right) g_t,
\]
where \( g_t \) is the gradient at time step \( t \). The second moment estimate \( v_t \) is updated as 
\[
v_t = \beta_2 v_{t-1} + \left(1 - \beta_2\right) g_t^2.
\]
The key difference in AMSGrad is the update for the maximum of the second moment estimate:
\vspace{-1em}
\[
\widehat{v_t} = \max\left({\widehat{v}}_{t-1}, v_t\right),
\]
which ensures that the second moment estimate does not decrease, thus avoiding the problem of vanishing step sizes. The parameter update rule is then given by
\[
\theta_{t+1} = \theta_t - \frac{\eta}{\sqrt{\widehat{v_t}} + \epsilon} m_t,
\]
where \( \theta_t \) represents the model parameters at step \( t \), \( \eta \) is the learning rate, and \( \epsilon \) is a small constant added to prevent division by zero.

The hyperparameters of the AMSGrad algorithm include the learning rate \( \eta \), which acts as a global scaling factor for the step size and may either be fixed or evolve during training with a scheduler. The decay rates \( \beta_1 \) and \( \beta_2 \) control the exponential decay of the first and second moment estimates, respectively. These values are typically set to \( \beta_1 = 0.9 \) and \( \beta_2 = 0.999 \). The small constant \( \epsilon \), typically set to \( 10^{-8} \), is used to avoid division by zero during rescaling.

In addition, the learning rate scheduler utilizes two important parameters: patience, which determines how many consecutive epochs without improvement in the validation loss will be tolerated before the learning rate is adjusted, and the decay factor, which determines how quickly the learning rate decreases during training. These features of AMSGrad make it a more stable optimizer compared to the standard Adam algorithm, contributing to more reliable convergence during training.

\newpage
\subsection*{S6. Residual plot for NequIP-BAM model and NACV prediction}

\begin{figure}[h]
\centering
\includegraphics[width=6in]{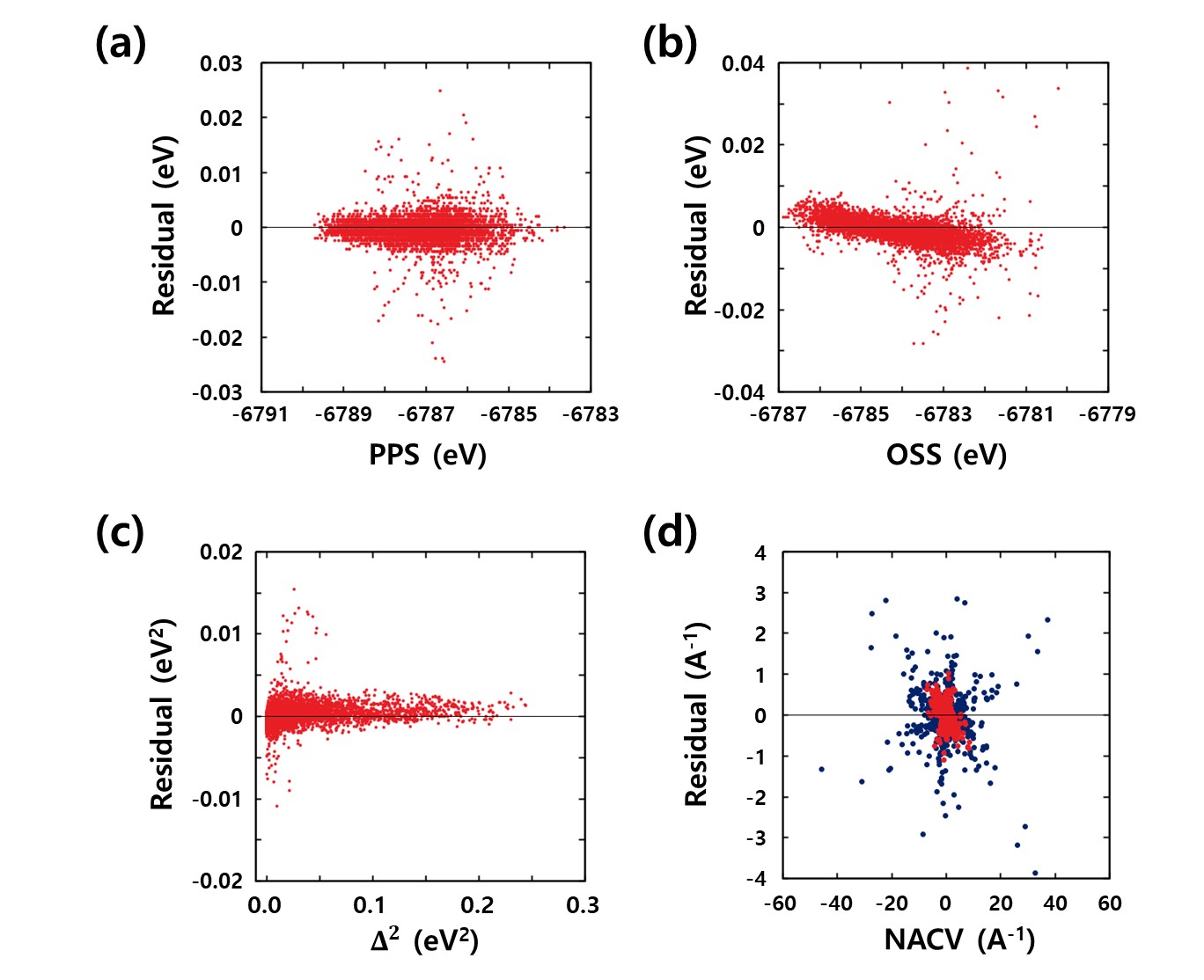}
\caption{Residual plots for the errors are presented for (a) PPS, (b) OSS, (c) $\Delta^2$ from the NequIP-BAM model, and (d) NACV elements calculated from the components of the diabatic Hamiltonian. In plot (d), the red and blue dots represent the residuals from regions far from and vicinity of the CI, respectively.} 
\label{fig:residual}
\end{figure}

\newpage
\subsection*{S7. Discontinuity in real-valued eigenvector coefficients around conical intersections}

\begin{figure}[h]
\centering
\includegraphics[width=6.5in]{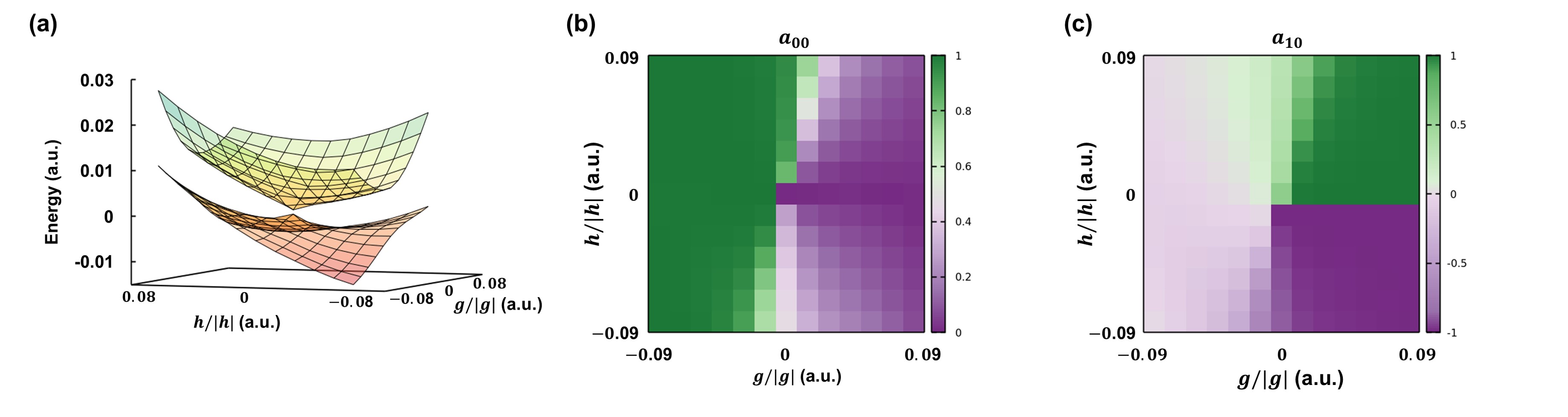}
\caption{(a) Adiabatic potential energy surfaces ($E_0^{SSR}$ and $E_1^{SSR}$) around the conical intersection, and discontinuous real-valued eigenvector coefficients (b) $a_{00}$ and (c) $a_{10}$ in the branching plane obtained from the ML diabatic Hamiltonian. By propagating the phase-aligned (adiabatically changing) eigenvector coefficients $a_{00}$ and $a_{10}$ corresponding to the lowest adiabatic state within the branching space, we can show that our approach can describe the Berry phase around a CI.} 
\label{eigvec}
\end{figure}


\newpage
\subsection*{S8. Energies and nonadiabatic coupling matrix elements (NACMEs) for sample trajectories}

\begin{figure}[h]
\centering
\includegraphics[width=6in]{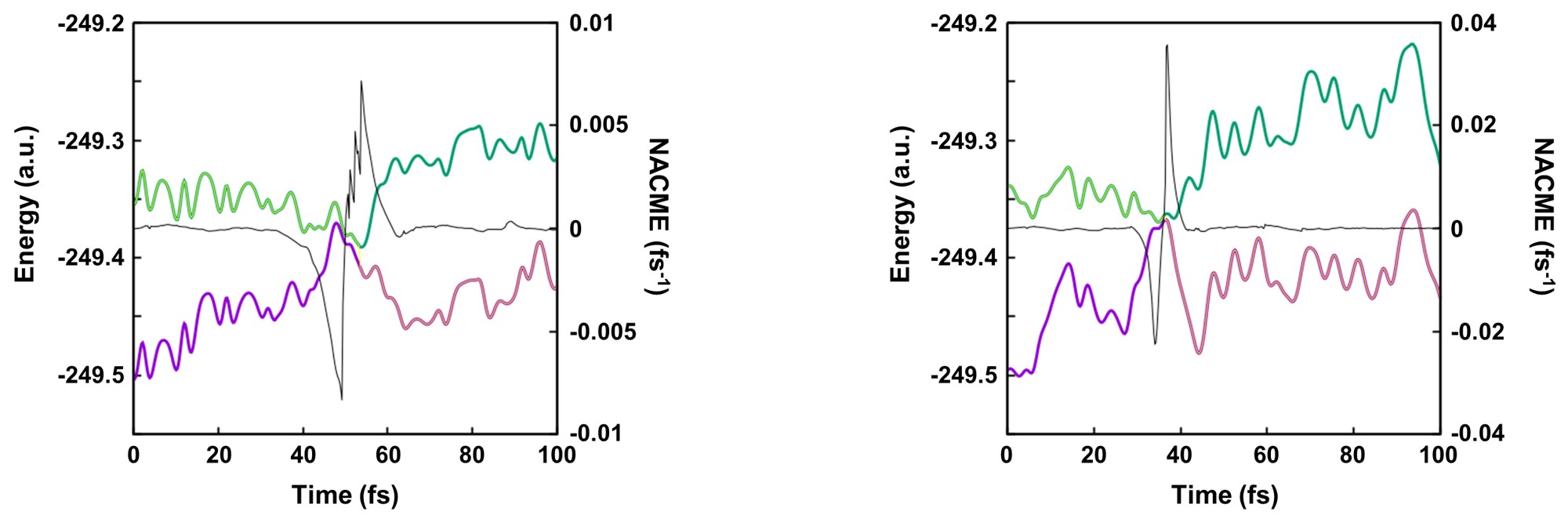}
\caption{Energies (purple: the ground state, green: the first excited state) and NACMEs (black) along two sampled trajectories from the ML-ESMD dynamics simulations. Energy of the active state in SHXF dynamics is depicted in yellow. The NACME between the ground and the first excited states, $\sigma_{01}$, is calculated as $\sigma_{01} = \Sigma_v d_{ij, v} \cdot \dot{\mathbf{R}}_v$ with the nuclear velocity $\dot{\mathbf{R}}_v$ of the $\nu$th nucleus.} 
\label{fig:individual_traj}
\end{figure}

\newpage
\subsection*{S9. XYZ structures of conical intersection and their branching plane vectors}

\begin{table}[h!]
    \centering
    \begin{tabular}{lccc}
        \hline 
        14 \\
        \multicolumn{4}{l}{CI structure}  \\
        C  & -2.8938847346  & -1.2816343136 &  -0.0975451573 \\
        C  & -1.5040815166  & -1.1810295518 &  -0.3268286915 \\
        H  & -3.6107042866  & -0.5296514365 &  -0.5519596694 \\
        H  & -3.3362279331  & -2.1779062651 &   0.3789458932 \\
        H  & -1.2095650879  & -1.9138801691 &  -1.1295700818 \\
        C  & -0.7311542226  & -0.1737470473 &   0.2503998578 \\
        C  &  0.6552761641  &  0.0181298130 &   0.0975004387 \\
        H  & -1.2820857283  &  0.4513001486 &   0.9485008590 \\
        H  &  1.2533183965  & -0.8775230227 &   0.1965584710 \\
        C  &  1.3560122740  &  1.2020645377 &  -0.0707564267 \\
        H  &  0.7557851866  &  2.0858939465 &  -0.4366375873 \\
        N  &  2.6880973064  &  1.3081777377 &   0.0915157683 \\
        H  &  3.2072727088  &  2.1621611590 &  -0.1102623665 \\
        H  &  3.2458481377  &  0.5062003077 &   0.4519509678 \\
        \hline 
    \end{tabular}
\end{table}

\begin{table}[h!]
    \centering
    \begin{tabular}{lccc}
        \hline 
        14 \\
        \multicolumn{4}{l}{Difference gradient vector ($\Vec{\boldsymbol{g}}$)}  \\
      C  &    -0.034369993   &  -0.005188318   &    0.005785784  \\
      C  &     0.028200140   &   0.006307650   &   -0.005728461  \\
      H  &     0.002842261   &  -0.011927020   &   -0.016648926  \\
      H  &     0.002516529   &   0.011871241   &    0.017034645  \\
      H  &    -0.000621411   &   0.016874578   &   -0.011201068  \\
      C  &    -0.003841267   &  -0.020100285   &    0.017426507  \\
      C  &    -0.001052722   &  -0.002570883   &   -0.000719742  \\
      H  &     0.002332728   &  -0.001647303   &   -0.001171673  \\
      H  &     0.000696259   &  -0.000366154   &    0.000459735  \\
      C  &     0.008602508   &   0.008488679   &   -0.006380110  \\
      H  &    -0.000407865   &  -0.000346202   &   -0.000646319  \\
      N  &    -0.006243598   &  -0.001851112   &    0.000745113  \\
      H  &     0.000337367   &   0.001484771   &    0.000798723  \\
      H  &     0.001009065   &  -0.001029643   &    0.000245793  \\
        \hline 
    \end{tabular}
\end{table}

\begin{table}[h!]
    \centering
    \begin{tabular}{lccc}
        \hline 
        14 \\
        \multicolumn{4}{l}{Coupling derivative vector ($\Vec{\boldsymbol{h}}$)}  \\        
       C  &   -0.004364314   &    0.002200454  &    0.002603525 \\
       C  &    0.001946064   &   -0.000413203  &   -0.001837264 \\
       H  &    0.001442373   &    0.001896739  &    0.004333536 \\
       H  &    0.001164312   &   -0.001857010  &   -0.004913866 \\
       H  &    0.002804150   &   -0.003183688  &    0.003316332 \\
       C  &   -0.001616837   &    0.001346462  &   -0.004304842 \\
       C  &   -0.002186741   &   -0.003386407  &    0.001259826 \\
       H  &   -0.000632464   &    0.000784265  &    0.000065353 \\
       H  &    0.000158170   &   -0.000055186  &   -0.000804227 \\
       C  &    0.003650495   &    0.003288490  &    0.000966926 \\
       H  &   -0.000015796   &   -0.000037253  &   -0.000114570 \\
       N  &   -0.002789839   &   -0.000620654  &   -0.000560910 \\
       H  &    0.000189445   &    0.000366726  &   -0.000182004 \\
       H  &    0.000250982   &   -0.000329735  &    0.000172185 \\
        \hline 
    \end{tabular}
\end{table}

\clearpage
\bibliography{achemso}